\documentclass[a4paper,11pt]{article}
\pdfoutput=1 

\usepackage{jinstpub} 
                                   
\usepackage{epsfig}
\usepackage{graphicx}                  
\usepackage{ae}
\usepackage{amsmath}
\usepackage{amssymb}
\usepackage{graphics}
\usepackage{dcolumn}
\usepackage{bm}

\title{\boldmath Characterisation of an RPC prototype with moderate resistivity plates using 
 tetrafluoroethane ($C_2H_2F_4$) }

\author[a, 1]{Arindam Sen,\note{Corresponding author.}}
\author[a]{Sayak Chatterjee,}
\author[a]{Shreya Roy}
\author[a, 2]{Saikat Biswas\note{E-mail: saikat@jcbose.ac.in}}
\author[a]{Supriya Das}

\affiliation[a]{Department of Physics and Centre for Astroparticle Physics and Space Science (CAPSS), \\Bose Institute, EN-80, Sector-V, Bidhannagar, Kolkata-700091, India}

\emailAdd{arindam@jcbose.ac.in}

\abstract{Keeping in mind the requirements of high rate capable, cost effective, large area detectors to be used 
in future high energy physics experiments, commercially available bakelite plates having moderate bulk resistivity are used to build an RPC module. The chamber is tested with cosmic rays in the avalanche mode using 100\% Tetrafluoroethane ($C_2H_2F_4$). Standard NIM electronics are used for this study. The efficiency, noise rate and time resolution are measured. The detailed method of measurement and the first test results are presented.
}

\keywords{Gaseous detectors; Gaseous imaging and tracking detectors; Resistive-plate chambers}

\arxivnumber{2004.05469} 

\proceeding{XV$^{\text{th}}$ Workshop on Resistive Plate Chambers and Related Detectors - RPC2020\\
  10-14 February, 2020\\
  Rome, Italy}

\begin{document}
\maketitle
\flushbottom

\vspace{-0.4cm} 
\section{Introduction}
\label{sec:intro}
Since the invention of the Resistive Plate Chambers (RPCs) \cite{Santonico}  as a cost effective technology that can be used to 
build large area granular, reasonably fast and high rate capable detectors, it has found use not only in a 
large number of high energy physics experiments \cite{BABAR, STAR, ALICE, ATLAS, CMS, HADES} as trigger, Time of Flight (TOF) and tracking devices but
also in several cosmic ray experiments \cite{BESS, ARGO} and Neutrino experiments \cite{OPERA, DAYABAY, SB09}. Future experiments such as Compressed Baryonic Matter 
(CBM) at FAIR also propose to use RPCs as one of the key detectors \cite{CBM}.     

Keeping in mind the possibility of using RPCs as future high rate ($\sim$~15~kHz/cm$^2$) capable tracking detectors, we have taken up a study to characterise an RPC prototype built using a particular type of bakelite plates with moderate bulk resistivity.
The prototype is tested with 100\% Tetrafluoroethane ($C_2H_2F_4$) gas for the first time in this work.


\section{Detector description and experimental set-up}
\label{sec:setup}

The detector prototype is built with two 2~mm thick bakelite plates, each having dimension 30~cm~$\times$~30~cm and bulk resistivity $3~\times~10^{10}~\Omega~cm$ (at 22$^\circ$C temperature and 60\% Relative humidity) and without any linseed oil coating inside. The gas gap is maintained with four edge spacers having width 1~cm, thickness 2~mm and one button spacer having 1~cm diameter, thickness 2~mm. Both the spacers are made of perspex (resistivity $\sim$~$10^{15}~\Omega~cm$). The surface resistivity of the graphite layer is measured to be $\sim$ 500~$k\Omega/\square$. Two 1~cm~$\times$~1~cm copper tapes are used at two diagonally opposite corners to apply high voltage (HV). HV of opposite polarities are applied on two sides. To collect the signals copper pick-up panels are used. They are made of 2.5~cm wide strips with a separation of 2~mm between two consecutive ones.

The signals from the pick-up strips are fed to a 10X fast amplifier and then to the discriminator. The cosmic ray master trigger is made using three fast plastic scintillators. Among them, two scintillators (with dimensions 10~cm~$\times$~10~cm and 2~cm~$\times$~10~cm respectively) are placed above and one (with dimension 20~cm~$\times$~20~cm) is placed below the RPC module. The scintillators make the trigger window of area 2~cm~$\times$~10~cm. Thresholds to the discriminators are set to -15~mV for all the scintillators and also for the RPC. The width of the 3-Fold scintillator master trigger is set to 150~ns. Finally, the discriminated RPC signal from one single strip is taken in coincidence with the 3-Fold master trigger and a 4-Fold NIM signal is obtained. The ratio of the 4-Fold signal and the 3-Fold scintillator signal is defined as the efficiency of the detector. The single RPC signals are also counted for a particular duration and the rate is defined as the noise rate of the chamber.

\begin{figure}[htbp]
\centering 
\includegraphics[scale=0.5]{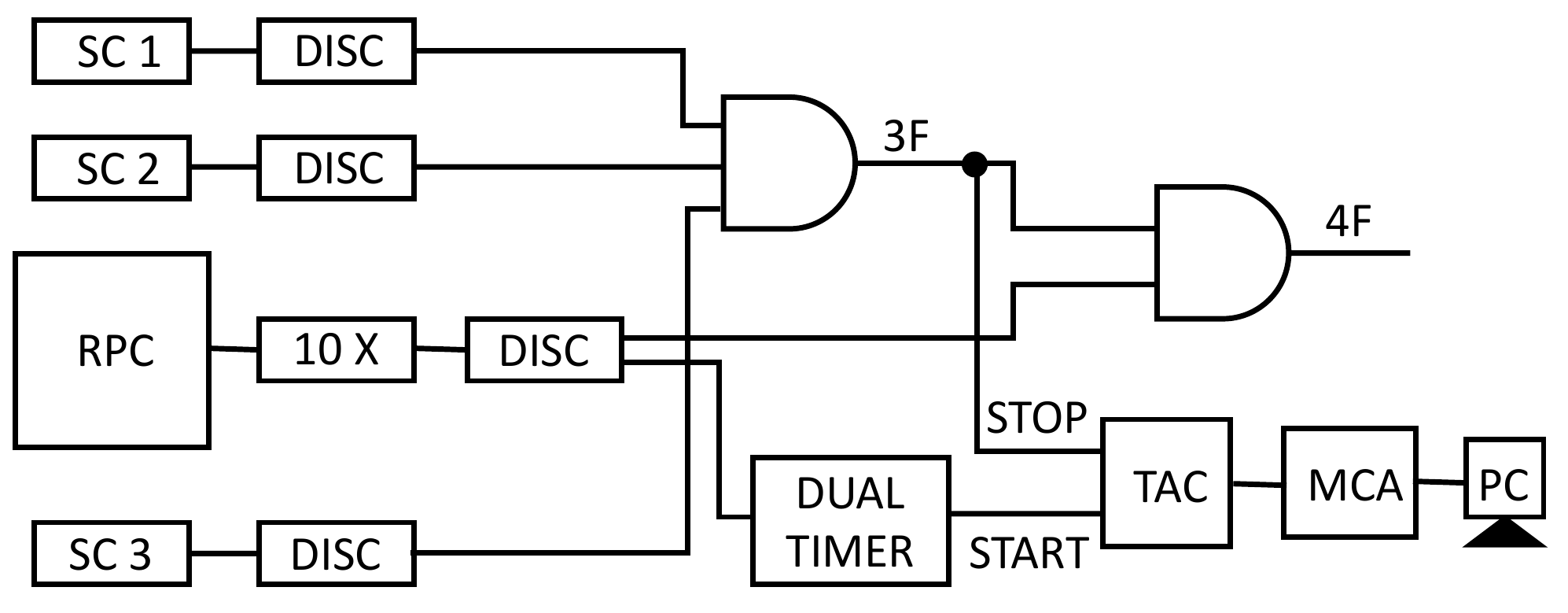}
\vspace{-0.4cm} 
\caption{\label{circuit} Schematic representation of the cosmic ray test setup. SC 1, SC 2 and SC 3 are the plastic scintillators of dimensions 10~cm~$\times$~10~cm, 2~cm~$\times$~10~cm and 20~cm~$\times$~20~cm. respectively. DISC, 10 X, TAC, MCA and PC are the discriminators, 10X fast amplifier, Time to Amplitude Converter, Multi Channel Analyser and Personal Computer respectively.}
\end{figure}

To measure the timing properties of the RPC, the same set-up is used. The discriminated RPC signal is stretched by a dual timer and fed as the START signal of the Time to Amplitude Converter (TAC). The 3-Fold scintillator coincidence signal is taken as the STOP signal input of the TAC. The output of the TAC is fed to the Multi Channel Analyser (MCA) and the spectra are stored in a Personal Computer (PC). Fig.~\ref{circuit} shows the schematic of the set-up for testing the RPC module using cosmic rays.

During the whole measurement, the temperature and relative humidity inside the laboratory are maintained at $\sim$~18-20$^\circ$C and 37-40\% respectively whereas the atmospheric pressure is monitored to be 1009-1020~mbar.

\vspace{-0.01cm}

\section{Results}
\label{sec:results}

\begin{figure}[htbp]
\centering 
\includegraphics[scale=0.55]{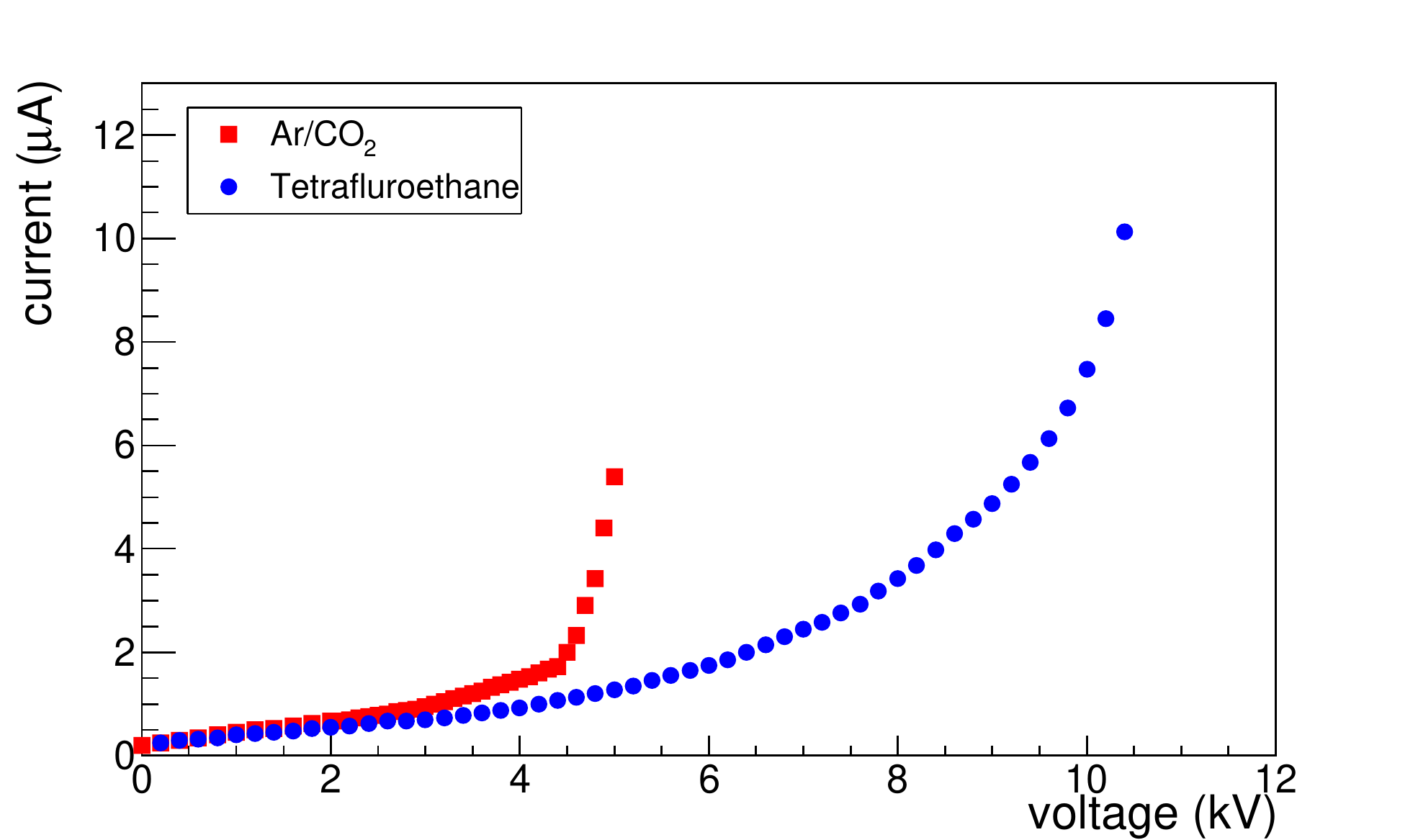}
\vspace{-0.4cm} 
\caption{\label{iv} The I-V Characteristics of RPC with two gas mixtures.}
\end{figure}

In this work, the efficiency, noise rate, time difference of RPC signal and the master trigger and time resolution of an oil-less bakelite RPC as a function of voltage are measured with cosmic rays. The detector current as a function of the bias voltage is shown in Fig.~\ref{iv}. It is visible that initially the current increases slowly with the voltage and above 8~kV across the gas gap the increase becomes rapid. At 8~kV voltage difference across the gap the signal of amplitude $\sim$~10-15 mV is observed in the oscilloscope at 50~$\Omega$ termination. The result has been compared with the one obtained using Argon/CO$_2$ gas in 70/30 ratio. Sharp breakdown in the I-V characteristics resulted with Argon/CO$_2$ at a lower voltage compared to that with the Tetrafluoroethane.

\begin{figure}[htbp]
\centering 
\includegraphics[scale=0.55]{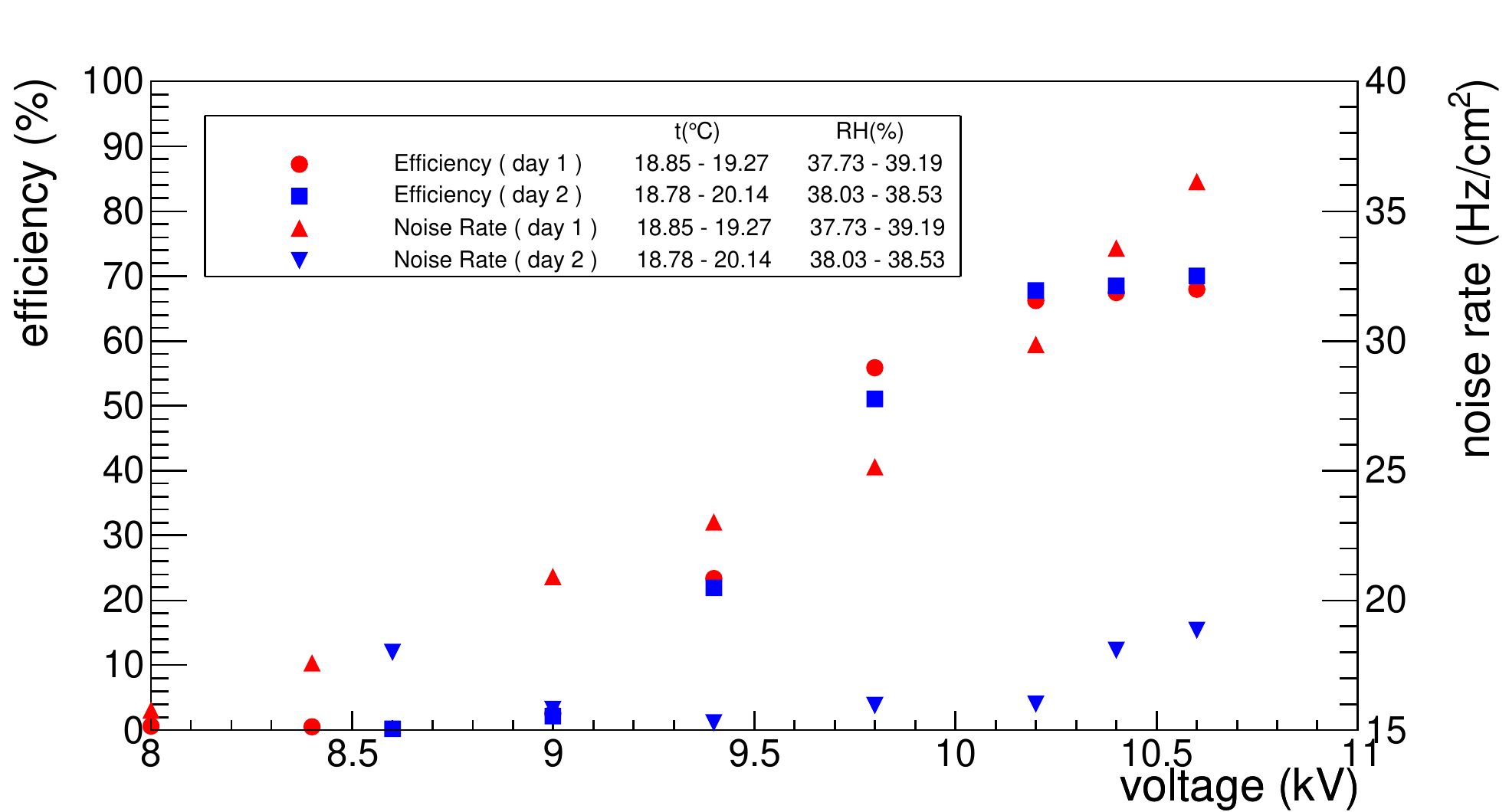}
\vspace{-0.4cm} 
\caption{\label{efficiency} Efficiency and noise rate as a function of voltage.}
\end{figure}

The noise rate as a function of voltage is measured for two consecutive days keeping the discriminator threshold at -15 mV and the results are  shown in Fig.~\ref{efficiency}. It is seen that for both days the noise rate increases with applied voltage but on the second day, the noise rate is found to be much less than that of the first day, because of better conditioning. The conditioning is done with continuous gas flow and keeping 4~kV across the gas gap over-night. It is to be mentioned here that, for linseed oil treated RPCs the noise rate with cosmic rays is found to be one order of magnitude better in some cases \cite{ATLAS08} and even two orders of magnitude for streamer operated RPCs with higher resistivity \cite{Opera06, Opera05, alice12}.

From Fig.~\ref{efficiency} it can also be seen that the efficiency starts increasing from 9~kV and saturates at a value of 70\% from 10.2~kV onwards. The same result is observed on both days. 

\begin{figure}[htbp]
\centering 
\includegraphics[scale=0.52]{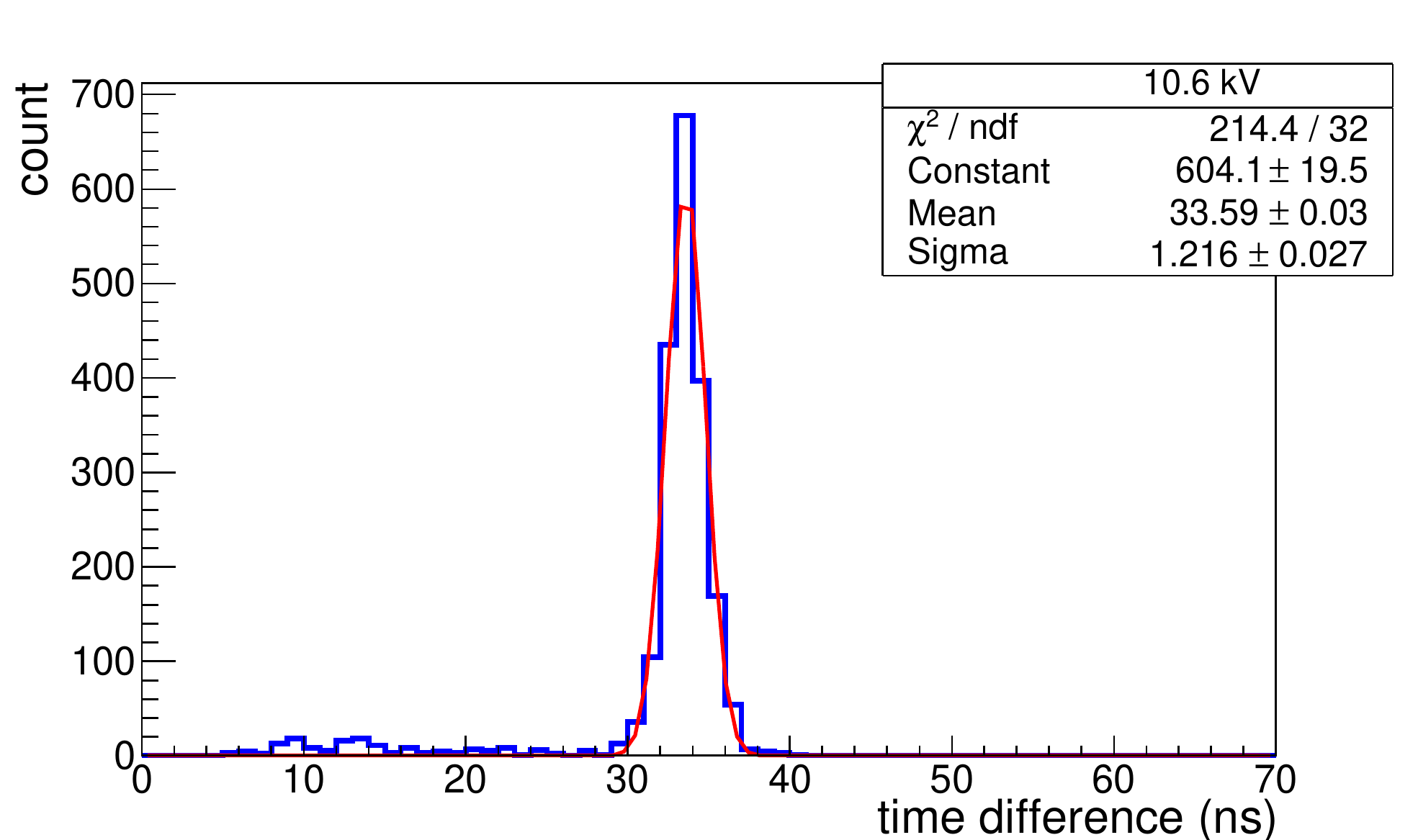}
\vspace{-0.4cm} 
\caption{\label{spectrum} Time spectrum of RPC at a voltage difference 10.6~kV across the gas gap.}
\end{figure}

While measuring the time resolution the RPC signal is stretched to 500~ns to avoid the effect of double or reflection pulses if there is any. The full scale of the TAC is set to 100~ns. The typical time spectrum for the RPC  is shown in Fig.~\ref{spectrum}. The distribution of the time difference between the RPC signal and the master trigger is fitted with the Gaussian function. Finding the $\sigma$ of the distribution and subtracting the contribution from the scintillator in quadrature the intrinsic time resolution of the RPC is calculated.

\begin{figure}[htbp]
\centering 
\includegraphics[scale=0.55]{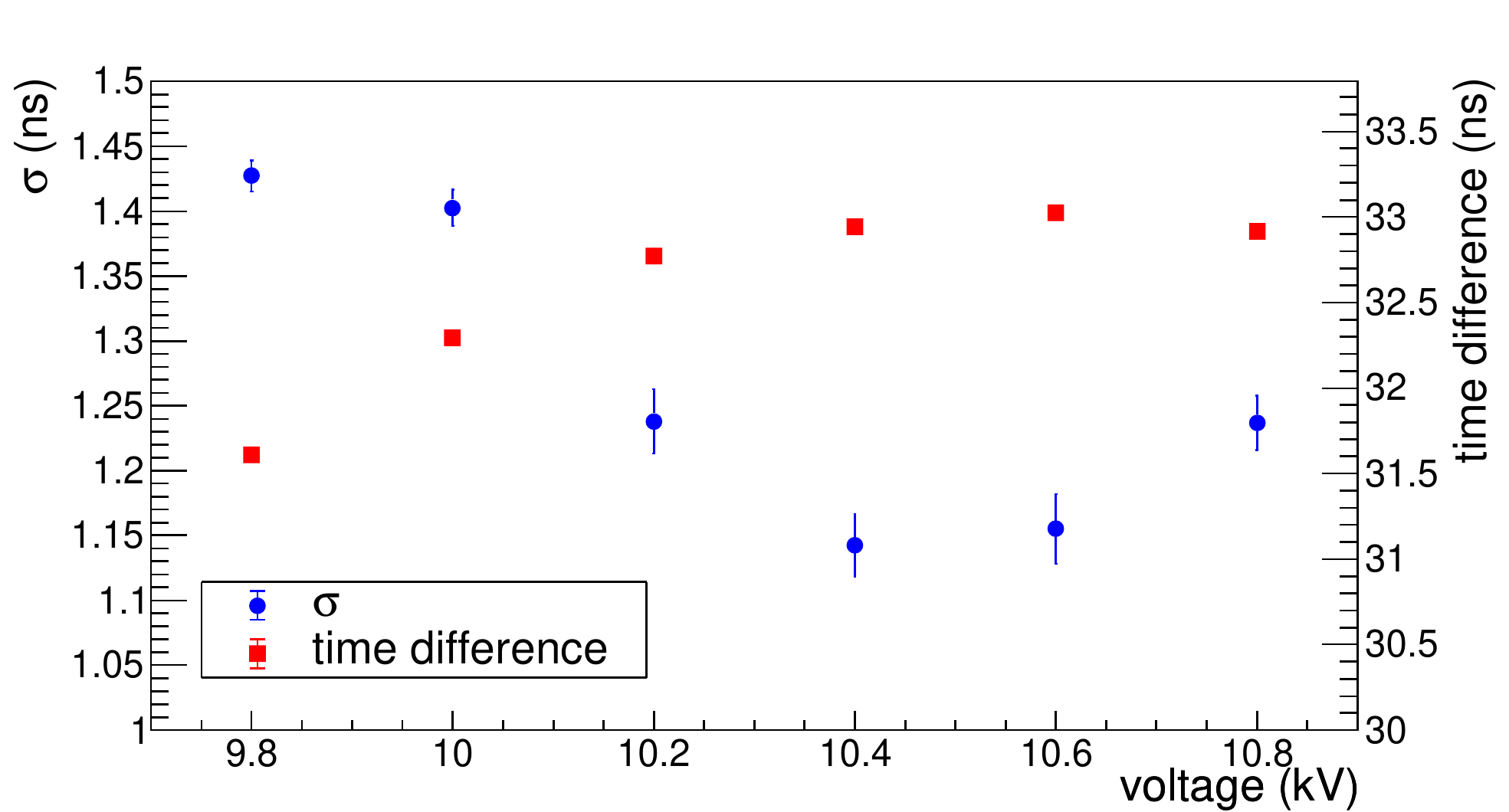}
\vspace{-0.4cm} 
\caption{\label{time_reso} Time resolution and time difference of RPC signal and master trigger as a function of voltage.}
\end{figure}

The time difference of the RPC signal with respect to the master trigger and the time resolution ($\sigma$) of the RPC as a function of the applied voltage are shown in Fig.~\ref{time_reso}. Since the RPC signal is used as the START signal and with the increase of the applied voltage the electric field inside the RPC becomes stronger, electrons travel faster and the signals arrive earlier. As a result with the increase of applied voltage the time difference increases and reaches a plateau from 10.2~kV onwards. The time resolution ($\sigma$) decreases and a value $\sim$~1.2~ns is obtained from 10.2~kV.

\vspace{-0.2cm} 
\section{Summary and Outlook}
\label{sec:summary}
An oil-less single gap RPC prototype is built with indigenous bakelite plates having bulk resistivity $3~\times~10^{10}$ $\Omega~cm$. The chamber is tested in the avalanche mode with 100\% Tetrafluoroethane gas. With this prototype, an efficiency $\sim$~70\% and time resolution 1.2~ns ($\sigma$) are obtained from an applied voltage of 10.2~kV onwards. Investigation of the reason behind lower efficiency is going on. One probable reason for the limitation in the efficiency is the voltage drop on the electrodes because of high current. Other Tetrafluoroethane based conventional gas mixtures will be tried. Estimation of the induced signal charge and the long-term stability test of this particular chamber is also in future plan.

High rate handling capability is one of the crucial factors for detectors to be used in many current and future high energy physics experiments. In that direction, we are searching for indigenous bakelite plates with better surface smoothness and lower resistivity. 

\vspace{-0.15cm} 
\acknowledgments

The authors would like to thank Prof. Sanjay K. Ghosh, Prof. Sibaji Raha, Prof. Rajarshi Ray and Dr. Sidharth K. Prasad for valuable discussions and suggestions in the course of the study. We would also like to thank Ms. Aayushi Paul of University of Calcutta and Ms. Rituparna Banerjee of IIT-ISM, Dhanbad for the calibration of the TAC and measurement of bulk resistivity of bakelite plates. This work is partially supported by the research grant SR/MF/PS-01/2014-BI from DST, Govt. of India and the research grant of CBM-MuCh project from BI-IFCC, DST, Govt. of India. A. Sen acknowledges his Inspire Fellowship research grant [DST/INSPIRE Fellowship/2018/IF180361]. S. Biswas acknowledges the support of Intramural Research Grant provided by Bose Institute.



\end{document}